\documentclass[aps,prb,twocolumn,groupedaddress]{revtex4-2}

\usepackage{amsmath}
\usepackage{graphicx}

\begin{document}


\title{Generation of non-equilibrium spin-motive force at the surface of a topological insulator induced by current-driven  domain wall motion in the magnetic substrate }


\author{Reza Fazeli Mehrabani}
\affiliation{Department of Physics, Azarbaijan Shahid Madani University, 53714-161 Tabriz, Iran}
\author{Babak Abdollahipour}
\affiliation{Faculty of Physics, University of Tabriz, 51666-16471 Tabriz, Iran}
\author{Hakimeh Mohammadpour}
\affiliation{Department of Physics, Azarbaijan Shahid Madani University, 53714-161 Tabriz, Iran\\
	Condensed Matter Computational Research Lab, Azarbaijan Shahid Madani University, 53714-161, Tabriz, Iran}
\author{Arash Phirouznia}
\email[]{Phirouznia@azaruniv.ac.ir}
\affiliation{Department of Physics, Azarbaijan Shahid Madani University, 53714-161 Tabriz, Iran\\
Condensed Matter Computational Research Lab, Azarbaijan Shahid Madani University, 53714-161, Tabriz, Iran}
\date{\today}

\begin{abstract}
The generalized Faraday method can be used for computation of spin-motive force (SMF) within the unitary transformation approach. Meanwhile, the unitary transformation approach, which could be employed to obtain an effective collinear exchange interaction of non-collinear magnetic structures, cannot be used when the other parts of the Hamiltonian contain spin-dependent couplings. In the current study, the method of unitary transformation has been generalized to systems with spin-momentum coupling. In this way, the SMF at topological insulator surfaces has been computed analytically.
\end{abstract}

\maketitle
\section{Introduction}
The Berry phase represents an angular measure, varying from $0$ to $2\pi$, defining the comprehensive phase transformation of a state vector while moving along a trajectory in its parametric vector space. This phenomenon is also recognized as \textit{geometric phase} \cite{1}.
Berry's phase can also be defined in the context of electron dynamics within a periodic network of atomic states. The unique torus topology of the Brillouin zone reveals the presence of a nonzero Berry phase in this one-dimensional parametric space. In this case,  quantization of the Berry phase comes form the symmetry of Bloch wave functions within the Brillouin zone. By establishing a connection between Berry's phase and Wyckoff positions in the crystal within the band representations of space groups, it becomes possible to utilize Berry's phase for labeling energy bands in solids \cite{2}.
\\

Topological insulators are identified as a new class of materials that have unique electronic and magnetic properties. This type of materials have attracted significant attention in recent years \cite{3}. A topological insulator is a material with an interior that functions as an electrical insulator, while its surface acts as an electrical conductor.
This means that electrons can only move along the surface of the material  \cite{8}. Accordingly, the key feature of these materials is that electric current flows only on the surface or edges but not through the bulk of  material \cite{4}. This is why they are called topological insulators.
Some examples of materials that exhibit topological insulator properties \cite{5} include: Two-dimensional crystals like  boron nitride \cite{6}, Bismuth and antimony semimetal alloys \cite{3} and Carbon nanotubes \cite{3,4,5,6}.
Due to these unique properties, topological insulators can find applications in electronic devices and optical tools \cite{7}. For instance, they can be used in quantum computing devices, display tools, light-emitting diodes and lasers, and quantum computers \cite{6}.
\\

Spin-motive force (SMF) arises from magnetization dynamics, exerting a spin-dependent force on conduction electrons. Analyzing its impact on magnetization dynamics necessitates considering factors such as spin accumulation, spin diffusion, and spin-flip scattering. This is crucial due to the nonuniform nature of the spin motive force \cite{9}.
The SMF is influenced by the exchange interaction that couples the spin of conduction electrons with local magnetization, establishing a common foundation with spin-transfer torque (STT) \cite{10,11}. 
While STT handles the exchange of angular momentum between spin current and magnetization, SMF facilitates the transfer of energy within the interacting subsystems.
In the case of the topological insulators a nontrivial interplay between topological electronic states and magnetic dynamics can be expected \cite{12}.
\\

The dynamics of magnetization in nonuniform magnetic structures creates a spin-dependent force on conduction electrons through exchange coupling. This results in a spin-dependent electric field, known as SMF, manifests in various magnetic configurations, including magnetic domain walls, vortices, and staggered ferrimagnetic structures. The magnitude and direction of the SMF are influenced by the spatiotemporal derivative of magnetic structures, which, in turn, is dictated by the magnetic interaction and geometry of the sample.
\\
As described above exchange interaction of electrons with a moving domain wall is responsible for generation of the SMF. Meanwhile, in the case of topological insulators (TIs) there is another interaction of spin-momentum type which plays a very significant role and has a key feature in this class of materials. Charge-spin inter-conversion could be arises as a result of spin-momentum coupling which can be introduced in a system via the spin-orbit couplings  \cite{20,21}. It can be inferred that the spin-charge inter-conversion, that manifests itself in the Rashba and Dresselhaus spin-orbit couplings is a consequence of the spin-momentum coupling \cite{22,23,24}. Therefore, spin-momentum coupling of TI materials, can be the source of spin-charge inter-conversion. 
This phenomenon highlights the intricate relationship between magnetic dynamics and spin-dependent effects on conduction electrons  \cite{13}.
\\
A new mechanism has also been introduced for generation of SMF using surface acoustic waves (SAW) via the spin-vorticity coupling (SVC) \cite{15}. It has been shown that by applying an external magnetic field, a direct spin current and a spin current with twice the frequency (second harmonic) can be created. While, these spin currents in the z direction have the property of non-reciprocity. It has also been estimated that the electric voltage resulting from these spin currents is measurable in polycrystalline nickel. These spin currents can be generated in any ferromagnet, if only a SAW device is available. Therefore, spin current can significantly expand the applications of spintronics. In addition, the present results may be important for strain-related spintronics, such as the piezospintronic effect, flexible spintronics, and mechanical control of spin-orbit interaction \cite{15,rezaei2018}.
\\
Meanwhile, the researches of this field delve into the analysis of the spin motive electric field using numerical techniques as well as experimental observations. Applications of the SMF, has been investigated particularly in ferromagnetic and ferrimagnetic materials, where the theoretical studies of the SMF, emphasizing its intricate connection with the topology of the magnetic structure. The experimental findings presented showcase the SMF induced by various magnetic samples, taking into account factors such as ferromagnetic coupling, dipole–dipole interaction, and antiferromagnetic coupling \cite{16}.
\\
Non-collinear magnetic structure induced SMF is generally calculated by the unitary transformation approach in which the non-collinear magnetic part of the Hamiltonian is converted to a collinear diagonalizable effective Hamiltonian. This method has also been employed in calculation of domain wall resistance \cite{LevyZhang,PRBPhirouznia,PRBFallahi,PRBMajidi}. However, in the presence of both spin-momentum and exchange interactions these two terms could not be transformed into a collinear effective system by a given unitary transformation. In the current study we have developed the unitary transformation approach for this type of systems by reducing the dimension of effective system. In this case, the direction normal to the electric field driven current has been eliminated, by an additional unitary transformation. Then it can be realized that a diagonalizable Hamiltonian can be obtained and the method of unitary transformation can be generalized to the systems containing spin-momentum coupling. 
\\
\section{Generalized Faraday's Law into the Berry Phase}
Faraday's Law,  $\varepsilon =-\frac{d\phi }{dt}$ which relates the electromotive force $\varepsilon$ to the $\phi$ magnetic flux, is adapted in ferromagnetic materials to account for nonconservative spin forces. These adjustments ensure energy conservation in evolving order parameters of itinerant ferromagnets \cite{17}. In this way, a modification has been suggested to Faraday's Law in which the nonconservative spin forces have been included using the generalized Faraday's law
\begin{eqnarray}
\varepsilon =-\frac{\hbar }{(-e)}\frac{d\gamma }{dt},   
\end{eqnarray}
where $\gamma$ is a spin average of the Berry phase in which both charge and spin of conduction electrons have been taken into account \cite{17}. The traditional model of an itinerant ferromagnet must expand to include the Berry phase. The induced potential energy, $\varphi _{s}^{\pm }$ , varies for majority and minority electrons, denoted by upper and lower signs. In simple cases, $\varphi _{s}^{\pm }$ is defined as $\mp \mu {{B}_{i}}(r) $,where $ {{B}_{i}}(r) $ represents the magnitude of position-dependent internal field. Notably, the force $-\vec{\nabla }\varphi _{s}^{\pm }$ observed in the Stern-Gerlach experiment is conservative and does not contribute to the electromotive force $\varepsilon $. The spin Berry phase is given by $\gamma _{s}^{\pm }=\frac{1}{2}\int{\vec{A}_{s}^{\pm }.dr}$. Where $\vec{A}_{s}^{\pm }$ is the Berry connection for majority and minority spins which contributes to the overall spin force through the following relation,
\begin{eqnarray}
	& f_{s}^{\pm }=-\frac{\hbar }{2}\frac{\partial \vec{A}_{s}^{\pm }}{\partial t}-{{\nabla }_{{\vec{r}}}}\varphi _{s}^{\pm }. 
\end{eqnarray}
A comprehensive theory describing current-induced ferromagnetic domain wall motion in nanowires was presented in 2005 \cite{18}. Using modified Landau-Lifshitz-Gilbert equations, a theory which captures angular momentum transfer from conduction electrons to the current driven domain wall can be achieved \cite{18}. A key finding was the prediction of uniform steady-state wall motion proportional to current density in the absence of pinning, indicating perfect angular momentum transfer. When incorporating extrinsic pinning, the theory predicts a linear velocity-current relationship above a critical depinning current, consistent with experiments. In some researches it has been provided fundamental insights into the spin-transfer torque mechanism where a good agreement between theoretical predictions and current-driven domain wall dynamics data has been achieved \cite{18}.
This study also examined the complex dynamics influencing current-induced domain wall motion in magnetic nanowires. It highlighted discrepancies between theory and experiments, underscoring the need for deeper understanding of underlying mechanisms. Factors like threshold currents for wall displacement, role of vortex walls, and accounting for microscopic details were explored \cite{19}. The depinning processes and interplay of spin-torque and pinning forces were discussed. Overall, the findings emphasized integrating theory and experiments for understanding current-driven magnetic domain wall motion comprehensively \cite{19}.
\begin{figure}[h!]
	\centering
	\includegraphics[width=0.45\textwidth]{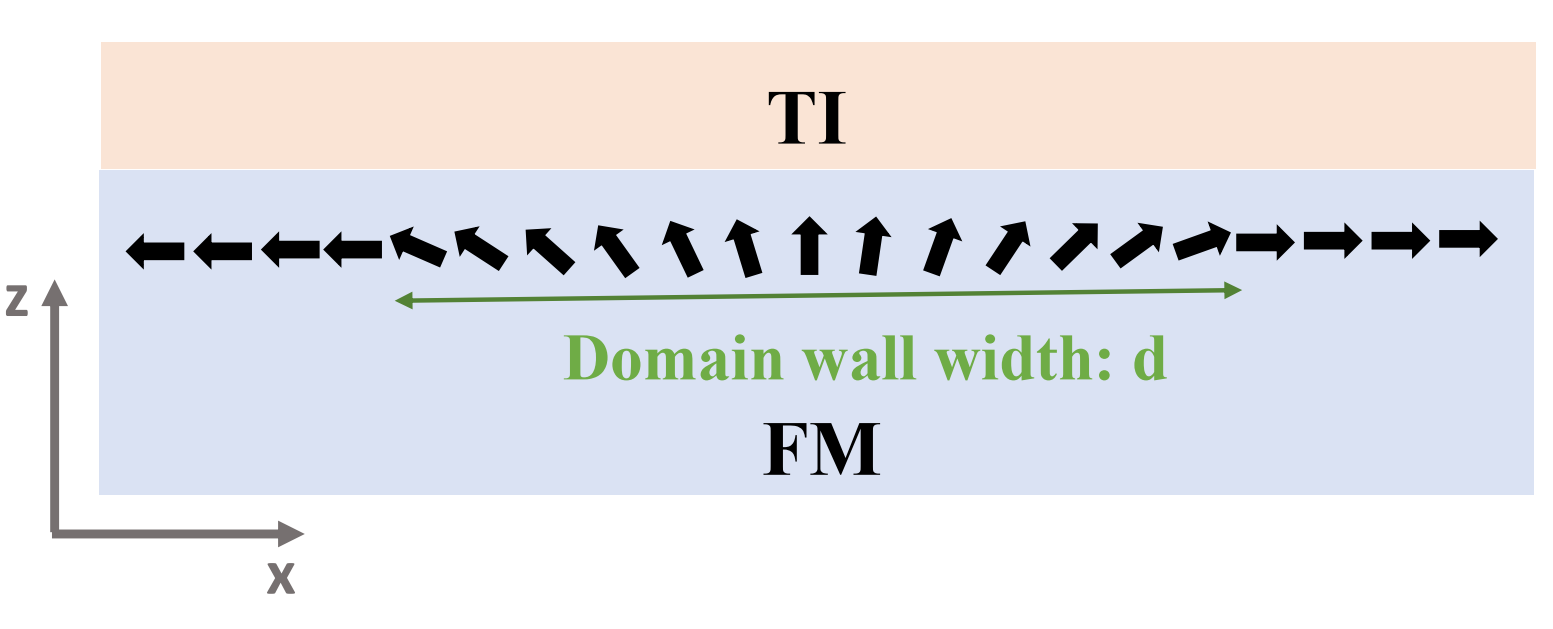} 
	\caption{Configuration of topological insulator and magnetic substrate.}
	\label{fig1}
\end{figure} 

\section{Method and Approach}
\begin{figure}[h!]
	\centering
	\includegraphics[width=0.5\textwidth]{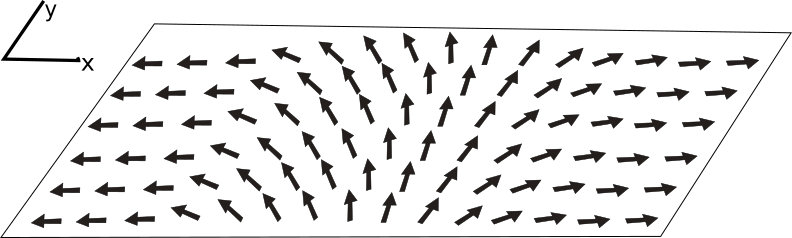} 
	\caption{Direction of the local magnetization in the substrate given by the assumed domain wall configuration. $y$ is the direction of rotation axis and local spins inside the domain wall are varying along the $x$ axis.}
	\label{fig2}
\end{figure}

\begin{figure}[h!]
	\centering
	\includegraphics[width=0.5\textwidth]{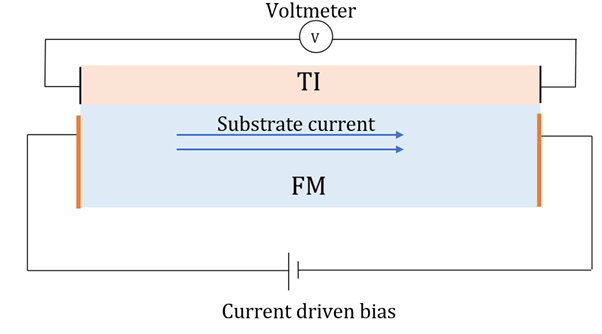} 
	\caption{Schematic view of SMF generated by substrate electric current. Substrate current results in domain wall motion and magnetization dynamics.  }
		\label{fig3}
\end{figure}

We consider the Hamiltonian of the topological insulator surface as follows:
\begin{eqnarray}
	& H={{v}_{F}}{{({{\vec{P}}}\times \sigma )}_{z}}+V(\vec{r})-JM\vec {\sigma .\hat{n}},
\end{eqnarray}
where ${{v}_{F}}$  is the Fermi velocity, $\sigma $ is the Pauli matrix, $V(r)$ is the system potential, $M$ is the induced magnetization of the ferromagnetic (FM) substrate at the topological insulator surface, $s$ is the spin of the conduction electrons, and $J$ represents the exchange interaction strength between the spin of conduction electron inside the topological insulator surface and the magnetic substrate. $n$ denotes the direction of local magnetization induced by substrate, which makes an angle $\theta $ with respect to the x-axis, in which the spatial dependence of this angle for a magnetic domain wall can be considered as $\theta (x,t)=\arccos (\tanh (x-{{x}_{0}}(t)/d))$ where ${x}_{0}$ is the center of the domain wall and $d$ denotes the domain wall width as shown in Fig. \ref{fig1}. It was assumed that the substrate contains a Neel type domain wall in which its rotation axis is the y-axis that was shown in Fig \ref{fig2}. \\
It was assumed that current driven domain wall motion in the underlying layer, is responsible for the magnetization dynamics in the system that was schematically shown in Fig \ref{fig3}. It can be realized that the magnetic field which is used to generate magnetic structure dynamics, by changing the spin configuration of local magnetization, cloud also change the majority and minority spins' population of conduction electrons. This may also result in secondary effects via the conduction and local spins interaction. Accordingly, the use of electric field have been proposed the to bring about magnetic dynamics. This approach could enable more efficient manipulation of magnetic states. Additionally, electric fields may offer finer control compared to magnetic fields alone. Future studies could explore the interplay between electric and magnetic fields. By using an electric field, for local spin dynamics we can selectively manipulate spin interactions providing a clearer understanding of spin dynamics. 

Using the following transformation operator:
\begin{eqnarray}
	& U={{e}^{i\frac{{{\sigma }_{y}}}{2}\theta }},
\end{eqnarray}
gives
\begin{eqnarray}
	& {\psi }'=U\psi, 
\end{eqnarray}
then the original Schrödinger equation 
\begin{eqnarray}
& H{\psi }=i\hbar \frac{\partial }{\partial t}{\psi },
\end{eqnarray}
can be converted to:
\begin{eqnarray}
\begin{aligned}
	 UH{{U}^{-1}}{\psi }'&=i\hbar U\frac{\partial }{\partial t}({{U}^{-1}}{\psi }') \\ 
	& =i\hbar U(\frac{\partial {{U}^{-1}}}{\partial t}){\psi }'+i\hbar U{{U}^{-1}}\frac{\partial {\psi'}}{\partial t} \\ 
	& =i\hbar U{{U}^{-1}}\frac{i{{\sigma }_{y}}}{2}\frac{\partial \theta }{\partial t}{\psi }'+i\hbar \frac{\partial {\psi }'}{\partial t}, \\ 
\end{aligned}
\end{eqnarray}
\begin{widetext}
then one can write
\begin{eqnarray}
\begin{aligned}
	 UH{{U}^{-1}}{\psi }'&=({{v}_{F}}U({{P}_{x}}{{\sigma }_{y}}-{{P}_{y}}{{\sigma }_{x}}){{U}^{-1}}  +JMU\sigma .\hat{n}{{U}^{-1}}){\psi }'\\
	&=-\hbar \frac{{{\sigma }_{y}}}{2}\frac{\partial \theta }{\partial t}{\psi }'+i\hbar \frac{\partial {\psi }'}{\partial t} \\ 
\end{aligned}
	\end{eqnarray}

\begin{eqnarray}
	\begin{aligned}
		 {{v}_{F}}(({{e}^{i{{\sigma }_{y}}{\theta }/{2}}}{{P}_{x}}{{\sigma }_{y}}{{e}^{-i{{\sigma }_{y}}{\theta }/{2}}}-{{P}_{y}}{{e}^{i{{\sigma }_{y}}{\theta }/{2}}}{{\sigma }_{x}}{{e}^{-i{{\sigma }_{y}}{\theta }/{2}}})  +JM{{\sigma }_{z}}){\psi }'=-\frac{\hbar }{2}{{\sigma }_{y}}\frac{\partial \theta }{\partial t}{\psi }'+i\hbar \frac{\partial {\psi }'}{\partial t}. \\ 
	\end{aligned}
	\end{eqnarray}
The transformation operator, $U$, is considered such that it aligns the rotated spins along the z-axis direction within the domain wall at the specific point x. It can be realized that this transformation satisfies
\begin{eqnarray}
	& U\sigma .n{{U}^{-1}}={{\sigma }_{z}}.
\end{eqnarray}
In this case, the remaining parts of the Hamiltonian i.e. the spin-momentum coupling terms transform under the $U$ operator as follows (see appendix A):
\begin{eqnarray}
	\begin{aligned}
		 U{{P}_{y}}{{\sigma }_{x}}{{U}^{-1}}&={{P}_{y}}{{e}^{i\frac{{{\sigma }_{z}}}{2}\theta }}{{\sigma }_{x}}{{e}^{-i\frac{{{\sigma }_{z}}}{2}\theta }} \\ 
		& ={{P}_{y}}(\cos \theta {{\sigma }_{x}}-\sin \theta {{\sigma }_{z}}), \\ 
	\end{aligned}
\end{eqnarray}
and 
\begin{equation}
	\begin{aligned}
		{{e}^{i\frac{{{\sigma }_{y}}}{2}\theta }}{{P}_{x}}{{\sigma }_{y}}{{e}^{-i\frac{{{\sigma }_{y}}}{2}\theta }} &= {{e}^{i\frac{{{\sigma }_{y}}}{2}\theta }}{{P}_{x}}{{e}^{-i\frac{{{\sigma }_{y}}}{2}\theta }}{{\sigma }_{y}} \\ 
		&= (-\frac{\hbar }{2}{{\sigma }_{y}}\frac{\partial \theta }{\partial x}+{{P}_{x}}){{\sigma }_{y}}.
	\end{aligned}
\end{equation}
\\
Meanwhile, for smooth domain walls with large $d$, $\theta$ could be considered as a linear function of space which can be obtained using the Taylor expansion one can write $\theta(x,t) = \frac{\pi}{2}-\frac{x-x_0(t)}{d}+\frac{(x-x_0(t))^3}{3d^3}+...$.
Therefore, Schrodinger equation can be written as

	\begin{eqnarray}
		\begin{aligned}
			 \left({{v}_{F}}\frac{\hbar}{2d}+{{v}_{F}}{{P}_{x}}{{\sigma }_{y}}-{{P}_{y}}({{\sigma }_{x}}\cos \theta +{{\sigma }_{z}}\sin \theta ) -\frac{\hbar }{2}\frac{{{v}_{D}}}{d}{{\sigma }_{y}}+JM{{\sigma }_{z}} \right){\psi'}=i\hbar \frac{\partial {\psi }'}{\partial t} \\ 
		\end{aligned}.
			\end{eqnarray}
In which we have defined the current driven domain wall velocity as $v_D = dx_0(t)/dt$. 
\end{widetext}
Unlike the exchange interaction which has been converted to a collinear effective Hamiltonian, as it can be inferred from the third term of the above Hamiltonian, since $\theta$ is a space dependent function, $U$-transformation cannot convert the spin-momentum coupling term into a collinear diagonalizable operator, however we can reduce the transverse direction in the momentum space by a new transformation $U_\lambda$  that eliminates the ${{P}_{y}}$ dependent part of the Hamiltonian.  This transformation is defined as:
\begin{eqnarray}
	{{U}_{\lambda }}={{e}^{i\lambda ({{P}_{y}}y+y{{P}_{y}})}},
\end{eqnarray}
accordingly, the second transformation gives the new wave function as
\begin{eqnarray}
	{\psi }''={{e}^{i\lambda ({{P}_{y}}y+y{{P}_{y}})}}{\psi }'
\end{eqnarray}

\begin{widetext}
	
Therefore, the effective Schrödinger equation for transformed wavefunction is given by:

\begin{eqnarray}
	\begin{aligned}
		\left({{v}_{F}}\frac{\hbar}{2d}+{{v}_{F}}{{P}_{x}}{{\sigma }_{y}}-U_\lambda{{P}_{y}}U_\lambda^{-1}({{\sigma }_{x}}\cos \theta +{{\sigma }_{z}}\sin \theta ) -\frac{\hbar }{2}\frac{{{v}_{D}}}{d}{{\sigma }_{y}}+JM{{\sigma }_{z}} \right){\psi''}=i\hbar \frac{\partial {\psi }''}{\partial t}. \\ 
	\end{aligned}
\end{eqnarray}
Given that $U_\lambda P_y U_\lambda^{-1}=0$ in the limit $\lambda \to \infty $ (see appendix B) 
\begin{eqnarray}
	\left({{v}_{F}}\frac{\hbar }{2d}+{{v}_{F}}{{P}_{x}}{{\sigma }_{y}}-\frac{\hbar {{v}_{D}}}{2d}{{\sigma }_{y}}+JM{{\sigma }_{z}}\right){\psi }''=i\hbar \frac{\partial {\psi }''}{\partial t}.\nonumber\\
\end{eqnarray}
\end{widetext}
By ignoring the constant term we arrive at the following equation:
\begin{eqnarray}
\left({{v}_{F}}{{P}_{x}}{{\sigma }_{y}} -\frac{\hbar {{v}_{D}}}{2d}{{\sigma }_{y}}+JM{{\sigma }_{z}}\right){\psi }''=i\hbar \frac{\partial {\psi }''}{\partial t}.
\end{eqnarray}
Then the effective Hamiltonian of the topological insulator in the k-space is given as follows:
\begin{eqnarray}
\begin{aligned}
	& H={{v}_{F}}\hbar {{k}_{x}}{{\sigma }_{y}}-\frac{\hbar {{v}_{D}}}{2d}{{\sigma }_{y}}+JM{{\sigma }_{z}} \\ 
	& =\hbar {{{{k}'}}_{x}}{{\sigma }_{y}}+JM{{\sigma }_{z}}. \\ 
\end{aligned}
\end{eqnarray}
Where
\begin{eqnarray}
	k_{x}'=k_x-\frac{v_D}{2dv_F}.
\end{eqnarray}
The eigenfunctions of the effective Hamiltonian are given as:
\begin{equation}
	{{{\psi }''}_{{{k}_{x}}+}}=\left( \begin{matrix}
		{{\alpha }_{+}}  \\
		{{\beta }_{+}}  \\
	\end{matrix} \right),{{{\psi }''}_{{{k}_{x}}-}}=\left( \begin{matrix}
		{{\alpha }_{-}}  \\
		{{\beta }_{-}}  \\
	\end{matrix} \right).
\end{equation}
Where 
\begin{equation}
	\begin{aligned}
		& {{\alpha }_{\pm }}=\frac{\hbar {{v}_{F}}{	k_{x}'}}{\sqrt{{{\hbar }^{2}}{{v}^{2}}_{F}{{k_{x}'}^{2}}+{{(JM-{{E}_{\pm }})}^{2}}}} \\ 
		& {{\beta }_{\pm }}=\frac{-i(JM-{{E}_{\pm }})}{\sqrt{{{\hbar }^{2}}{{v}^{2}}_{F}{{k_{x}'}^{2}}+{{(JM-{{E}_{\pm }})}^{2}}}}. \\ 
	\end{aligned}
\end{equation}
The corresponding eigen-energies take the form: ${{E}_{k_x'}^{\pm }}=\pm \sqrt{{{J}^{2}}{{M}^{2}}+{{\hbar }^{2}}k_{x}'^{2}v_{F}^{2}}$.
\\

By applying the reversed transformations introduced above on eigen wave functions, the eigen states of the original Hamiltonian can be obtained in the following form:
\begin{equation}
	{{\psi }_{{{k}_{x}}\pm }}={{e}^{-i\frac{{{\sigma }_{y}}}{2}\theta }}{{e}^{-i\lambda ({{P}_{y}}y+y{{P}_{y}})}}{{{\psi }''}_{{{k}_{x}}\pm }}.
\end{equation}
Considering the definition of the Berry phase given by:
\begin{equation}
	{{\gamma }^{\pm }}=i\int\limits_{0}^{x}{<{{\psi }_{\pm }}|\frac{\partial }{\partial r}|{{\psi }_{\pm }}>\cdot d\vec{r}},
\end{equation}
the integral can be performed over a closed path including the domain wall along the system and a semicircular contour which can be considered outside of the system, where the integral over the semicircular path will not contribute at the limit of infinite semicircle since the wave function vanishes outside of the system at this limit, and the main contribution to the integral comes from paths through the system where there are spatial variations of the magnetization. Therefore, we have:
\begin{equation}
	\begin{aligned}
		& {{\gamma }^{\pm }}=i\int\limits_{0}^{L}{dx({{\alpha }^{*}}_{\pm },{{\beta }^{*}}_{\pm }){{e}^{i{{\sigma }_{y}}\frac{\theta }{2}}}\frac{\partial }{\partial x}}{{e}^{-i{{\sigma }_{y}}\frac{\theta }{2}}}\left( \begin{matrix}
			{{\alpha }^{\pm }}  \\
			{{\beta }^{\pm }}  \\
		\end{matrix} \right) \\ 
		& =i\int\limits_{0}^{L}{dx({{\alpha }^{*}}_{\pm },{{\beta }^{*}}_{\pm })}\frac{-i{{\sigma }_{y}}}{2}\left( \begin{matrix}
			{{\alpha }^{\pm }}  \\
			{{\beta }^{\pm }}  \\
		\end{matrix} \right)\frac{\partial \theta }{\partial x}. \\ 
	\end{aligned}
\end{equation}
Finally, the Berry phase is obtained as follows:
\begin{equation}
	{{\gamma }^{\pm }}=-\frac{{{i}^{2}}}{2}({{\alpha }^{*}}_{\pm },{{\beta }^{*}}_{\pm }){{\sigma }_{y}}\left( \begin{matrix}
		{{\alpha }^{\pm }}  \\
		{{\beta }^{\pm }}  \\
	\end{matrix} \right)[\theta (L,t)-\theta (0,t)]
\end{equation}
In which $i<{{\psi }_{\pm }}|\frac{\partial }{\partial r}|{{\psi }_{\pm }}>$ is the Berry connection $ {{A}^{\pm }}(\vec{r})$,  $L$ is the dimension of system along the $x$ direction. Meanwhile, it should be noted that since the magnetization is varying just along the x-axis we have ${{A}_{y}}^{\pm }=0$. Therefore calculation of the gauge invariant Berry phase in a closed path could be reduced to the integral of  ${{A}_{x}}^{\pm }$ along the x-axis.  \\

Additionally, ${{\varepsilon }^{\pm }}$,  the spin-dependent motive force, is calculated as follows:
\begin{equation}
	{{\varepsilon }^{\pm }}=-\frac{\hbar }{e}\frac{d{{\gamma }^{\pm }}}{dt},
\end{equation}
accordingly,
\begin{equation}
	\begin{aligned}
		{{\varepsilon }_{k_x'}^{\pm }}&=-\frac{\hbar }{2e}({{\alpha }^{*}}_{\pm },{{\beta }^{*}}_{\pm }){{\sigma }_{y}}\left( \begin{matrix}
			{{\alpha }^{\pm }}  \\  
			{{\beta }^{\pm }}  \\  
		\end{matrix} \right) \\  
		& \times \frac{\partial }{\partial t} \left(\theta (L,t)-\theta (0,t)\right). \\  
	\end{aligned}
\end{equation}

Furthermore, when the Fermi energy is within the upper band width the spin motive force is obtained by the contribution of this band and full occupied lower band cannot contribute in the spin motive force (see appendix C).
\\

Where as mentioned before, $L$ is the length of the system. Now, assuming that ${v_D}$ being the velocity of substrate magnetic domain wall, the motion of domain wall center can be considered with the relation $x_0(t)=v_Dt$.  Therefore, after performing some algebra and simplifications, for the upper band we have:
\begin{equation}
	{{\varepsilon }_{k'_x}^{+}}=\frac{{{\hbar }^{2}}{{v}_{F}}{{k'}_{x}}}{e}\frac{JM-{{E}_{+}}}{{{\hbar }^{2}}{{v}^{2}}_{F}{{k'_x}^{2}}+{{(JM-{{E}_{+}})}^{2}}}\frac{v_D}{d}sech(\frac{-{{v}_{D}}t}{d}).
\end{equation}
One can obtain the total spin motive force by integrating over all of the contributing states as given by
\begin{eqnarray}	
	\varepsilon _{tot}^{+}&=&\sum\limits_{{{k}_{x}'}}^{{}}{\varepsilon _{{{k}_{x}'}}^{+}}\\ 
	& =&\int{\frac{d{{k}_{x}'}}{2\pi }\frac{{{\hbar }^{2}}{{v}_{F}}{{k}_{x}'}}{e}\frac{(JM-{{E}^{+}})}{{{\hbar }^{2}}v_{F}^{2}k_{x}'^{2}{{(JM-{{E}^{+}})}^{2}}}\frac{{{v}_{D}}}{d}\sec {{h}}(\frac{-{{v}_{D}}t}{d})}.  \nonumber
\end{eqnarray}
Using the fact that $E^2 = \hbar^2v_F^2{k'}_x^{2}+J^2M^2$ it can be realized that
\begin{eqnarray}
 \varepsilon _{tot}^{+}&=&\int{\frac{d{{k_{x}'}}}{2\pi }\frac{{\hbar }^{2}v_{F}k_{x}'}{e}}\frac{(JM-{{E}_{+}})}{2E_{+}^{2}-2JM{{E}_{+}}}\sec h(-\frac{{v}_{D}t}{d})\frac{v_{D}}{d}  \nonumber\\ 
 &=&-\frac{1}{2\pi e}\sec h(-\frac{{v}_{D}t}{d})\int\limits_{0}^{{E}_{F}}{\frac{dE}{2}}. \nonumber \\ 
\end{eqnarray}

At the steady state limit,  $t\sim L/v_D$ SMF of system has the following final expression::
	\begin{eqnarray}
		\varepsilon _{tot}^{+}=\frac{{{E}_{F}}}{2\pi ed}(\frac{{{v}_{D}}}{{{v}_{F}}}){{e}^{-\frac{L}{d}}}.
	\end{eqnarray}
\section{Conclusion}
As mentioned, in the present work, the dynamics of a magnetic substrate, consisting of two magnetic domains separated by a domain wall, in generating the spin motive force on the surface of a topological insulator has been investigated. The unitary transformation method is employed to transform the non-collinear magnetic system into a collinear one. This transformation is analytically feasible for systems without spin-momentum coupling. However, in the presence of spin-momentum coupling, the transformation introduces some terms involving spin rotation in real space, and the transformed system becomes non-collinear and unsuitable for analytical diagonalization.

In this work, after converting the exchange interaction to an effective collinear system using the first unitary transformation, we remove the position-dependent spin component of the Hamiltonian that arises from the spin-momentum coupling by another transformation. This results in an equivalent Hamiltonian that can be analytically diagonalized, that facilitates calculation of the Berry phase. Our calculations reveal that the motion of the magnetic wall in the substrate can induce an effective spin motive force on the surface of the topological insulator.

We have generalized the unitary transformation approach for SMF calculations to systems with spin-momentum coupling. Furthermore, the proposed method can be extended to similar systems involving spin-momentum coupling. The results indicate that the generated SMF strongly depends on the dimensions of the domain wall and its velocity, which is driven by the electric current in the substrate. 

As shown in this work, SMF appears to depend on the system dimensions. However, in the limit \( \frac{L}{d} \gg 1 \), the SMF becomes insensitive to the system size and eventually vanishes as \( t \to \infty \). In finite-size systems, where \( L \sim d \), the system experiences SMF within a characteristic time scale of \( \frac{d}{v_D} \).
\\

Our findings demonstrate that a significant SMF can be generated in the non-equilibrium regime induced by the electric field driving the domain wall motion. The SMF is inversely proportional to the domain wall width and directly proportional to the wall velocity. Therefore, in terms of the local magnetic field, the SMF is more likely to appear in the non-adiabatic regime, where the local magnetic field changes rapidly as experienced by spin-dependent carriers. This condition is typically met in fast-moving, narrow domain walls. 
\\

However, size of the domain wall must be sufficiently large since the SMF is exponentially suppressed by a factor of \( e^{-\frac{L}{d}} \). For sharp domain walls, the generated SMF is negligible. The optimal condition for obtaining a significant SMF occurs when \( L \sim d \). Therefore, an entirely non-adiabatic regime, characterized by very short domain walls, reduces the SMF. Accordingly, the best case for measuring SMF is in fast-moving, long, or smooth domain walls, which can be classified as a partial adiabatic regime.

\appendix*
\begin{widetext}
\section{A}
\label{A}
Using the following transformation
${\psi }'=U\psi$
Schrodinger equation in the presence of exchange interactions can be written as 
\begin{equation}
	\left( {{v}_{F}}U({{P}_{x}}{{\sigma }_{y}}-{{P}_{y}}{{\sigma }_{x}}){{U}^{-1}}+JM{{\sigma }_{z}} \right){\psi }'=i\hbar \frac{\partial {\psi }'}{\partial t}.
\end{equation}
\\
In which we have employed the following relation
\begin{eqnarray}
	& U\sigma .n{{U}^{-1}}={{\sigma }_{z}}.
\end{eqnarray}
Considering the fact that rotation angle, $\theta$, is $(x,t)$ dependent we can evaluate the momentum operator ${P_x}$ under the rotation transformation as follows
\begin{eqnarray}
				{{e}^{i\frac{{{\sigma }_{y}}}{2}\theta }}{{P}_{x}}{{\sigma }_{y}}{{e}^{-i\frac{{{\sigma }_{y}}}{2}\theta }}&=&{{e}^{i\frac{{{\sigma }_{y}}}{2}\theta }}{{P}_{x}}{{e}^{-i\frac{{{\sigma }_{y}}}{2}\theta }}{{\sigma }_{y}} \nonumber\\ 
			& =&{{e}^{i\frac{{{\sigma }_{y}}}{2}\theta }}\left(\left[ {{P}_{x}},{{e}^{-i\frac{{{\sigma }_{y}}}{2}\theta }} \right]+{{e}^{-i\frac{{{\sigma }_{y}}}{2}\theta }}{{P}_{x}}\right){{\sigma }_{y}} \nonumber\\ 
			& =&{{e}^{i\frac{{{\sigma }_{y}}}{2}\theta }}\left(\frac{\hbar }{i}\frac{\partial }{\partial x}{{e}^{-i\frac{{{\sigma }_{y}}}{2}\theta }}+{{e}^{-i\frac{{{\sigma }_{y}}}{2}\theta }}{{P}_{x}}\right){{\sigma }_{y}} \nonumber\\ 
			& =&{{e}^{i\frac{{{\sigma }_{y}}}{2}\theta }}\left(\frac{\hbar }{i}(-\frac{i{{\sigma }_{y}}}{2}\frac{\partial \theta }{\partial x}){{e}^{-i\frac{{{\sigma }_{y}}}{2}\theta }}+{{e}^{-i\frac{{{\sigma }_{y}}}{2}\theta }}{{P}_{x}}\right){{\sigma }_{y}} \nonumber \\ 
			& =&\left(-\frac{\hbar }{2}{{\sigma }_{y}}\frac{\partial \theta }{\partial x}+{{P}_{x}}\right){{\sigma }_{y}}. 
\end{eqnarray}
Momentum operator of the $y$ direction commutes with the rotation operator so using the relation of ${{e}^{i\frac{{{\sigma }_{y}}}{2}\theta }}=(\cos \frac{\theta }{2}+i{{\sigma }_{y}}\sin \frac{\theta }{2})$ we have: 
\begin{eqnarray}
	   U{{P}_{y}}{{\sigma }_{x}}{{U}^{-1}}&=&U{{P}_{y}}{{U}^{-1}}U{{\sigma }_{x}}{{U}^{-1}}\nonumber\\&=&{{P}_{y}}{{e}^{i\frac{{{\sigma }_{y}}}{2}\theta }}{{\sigma }_{x}}{{e}^{-i\frac{{{\sigma }_{y}}}{2}\theta }} \nonumber\\ 
	& =&{{P}_{y}}\left((\cos \frac{\theta }{2}+i{{\sigma }_{y}}\sin \frac{\theta }{2}){{\sigma }_{x}}(\cos \frac{\theta }{2}-i{{\sigma }_{y}}\sin \frac{\theta }{2}\right)\nonumber \\ 
	& =&{{P}_{y}}\left({{\cos }^{2}}\frac{\theta }{2}{{\sigma }_{x}}-i\cos \frac{\theta }{2}\sin \frac{\theta }{2}i{{\sigma }_{z}}+i\sin \frac{\theta }{2}\cos \frac{\theta }{2}(-i{{\sigma }_{z}})+{{\sin }^{2}}\frac{\theta }{2}{{\sigma }_{y}}{{\sigma }_{x}}{{\sigma }_{y}}\right)\nonumber \\ 
	& =&{{P}_{y}}\left({{\cos }^{2}}\frac{\theta }{2}{{\sigma }_{x}}+\frac{1}{2}\sin \theta {{\sigma }_{z}}+\frac{1}{2}\sin \theta {{\sigma }_{z}}-{{\sin }^{2}}\frac{\theta }{2}{{\sigma }_{x}}\right)\nonumber \\ 
	& =&{{P}_{y}}\left(\cos \theta {{\sigma }_{x}}-\sin \theta {{\sigma }_{z}}\right).
\end{eqnarray}
Substituting the transformed operators into the original equation results in:
	\begin{equation}
		\left( {{v}_{F}}(-\frac{\hbar }{2}{{\sigma }_{y}}\frac{\partial \theta }{\partial x}+{{P}_{x}}{{\sigma }_{y}}-{{P}_{y}}(\cos \theta {{\sigma }_{x}}-\sin \theta {{\sigma }_{z}}))+JM{{\sigma }_{z}} \right){\psi }'=i\hbar \frac{\partial {\psi }'}{\partial t}.
	\end{equation}
\section{B}
Applying the second unitary transformation i.e: 
\begin{equation}
	U_{\lambda }={{e}^{i\lambda ({{P}_{y}}y+y{{P}_{y}})}}.
\end{equation}
And using the following transformed wave function in the Schrodinger equation
	\begin{equation}
	{\psi }''={{e}^{i\lambda ({{P}_{y}}y+y{{P}_{y}})}}{\psi }',
	\end{equation}
one can obtain:
		\begin{equation}
		\begin{aligned}
			& \left({{v}_{F}}U_{\lambda }((-\frac{\hbar }{2}{{\sigma }_{y}}\frac{\partial \theta }{\partial x}+{{P}_{x}}){{\sigma }_{y}})U{{_{\lambda }}^{-1}}-U_{\lambda }{{P}_{y}}U{{_{\lambda }}^{-1}}(\cos \theta {{\sigma }_{x}}-\sin \theta {{\sigma }_{z}}) +JM{{\sigma }_{z}}\right){\psi }''=i\hbar \frac{\partial {\psi }''}{\partial t}. \\ 
		\end{aligned}
			\end{equation}
We have used the fact that $U_\lambda$ commutes with $P_x,~\sigma_{x},~\sigma_{y}$ and $\sigma_{z}$ then we can write
\begin{eqnarray}
	U_\lambda P_y 	U_\lambda^{-1} &=& {{e}^{i\lambda ({{P}_{y}}y+y{{P}_{y}})}} P_y {{e}^{-i\lambda ({{P}_{y}}y+y{{P}_{y}})}} 
	\nonumber	\\
	&=&
	P_y+i\lambda(2i\hbar P_y) +\frac{(i\lambda )^2}{2!}(2i\hbar)^2 P_y+ \frac{(i\lambda )^3}{3!}(2i\hbar)^3 P_y+...
	\nonumber	\\
	&=& \left(1-2\hbar\lambda+\frac{(2\hbar\lambda)^2}{2!}-\frac{(2\hbar\lambda)^3}{3!}+...\right)P_y
	\nonumber	\\
	&=&e^{(-2\hbar\lambda)}P_y,
\end{eqnarray}
	\end{widetext}
where at the limit of large $\lambda$ we have $U_\lambda P_yU_\lambda^{-1}\sim 0$. So $P_y $ depending term could be considered very small. Then final form of the equation after all transformations can be given as:
\begin{equation}
		\left({{v}_{F}}({{P}_{x}}-\frac{\hbar }{2}\frac{\partial \theta }{\partial x}{{\sigma }_{y}}){{\sigma }_{y}}+JM{{\sigma }_{z}}\right){\psi }''=i\hbar \frac{\partial {\psi }''}{\partial t}
\end{equation}

\section{C}
Generally, the total spin motive force is given by the following relation when both of the bands contribute, 
\begin{eqnarray}
	{{E}_{SMF}}=p{{\varepsilon }_{tot}^{+}},
\end{eqnarray}
where ${\varepsilon }^{+}$ is the spin motive force of the upper band  
and polarization is defined as:
\begin{equation}
	p=\frac{{{\sigma }^{+}}-{{\sigma }^{-}}}{{{\sigma }^{+}}+{{\sigma }^{-}}}.
\end{equation}
In which ${\sigma }^{\pm}$ is the conductivity of $\pm$ bands, in the current case when the lower band is full and the Fermi level is inside the upper band width we have $p = 1$.
\bibliographystyle{apsrev4-2.bst}
\bibliography{ref.bib}
\end{document}